\documentclass[preprint,showpacs,preprintnumbers,amsmath,amssymb]{revtex4-1}
\usepackage{graphicx}
\usepackage{epsfig}
\usepackage{epstopdf}
\begin{document}

\title{Ferromagnetic interactions and martensitic transformation in Fe doped Ni-Mn-In shape memory alloys}
\author{D. N. Lobo},
\address{Department of Physics, Goa University, Taleigao Plateau, Goa 403 206 India}
\author{K. R. Priolkar} \email[corresponding author:]{krp@unigoa.ac.in}
\address{Department of Physics, Goa University, Taleigao Plateau, Goa 403 206 India}
\author{S. Emura} 
\address{Institute of Scientific and Industrial Research, Osaka University, 8-1 Mihogaoka, Ibaraki,Osaka 567-0047, Japan}
\author{A. K. Nigam}
\address{Tata Institute of Fundamental Research, Dr. Homi Bhabha Road, Colaba, Mumbai 400 005 India}

\date{\today}

\begin{abstract}
The structure, magnetic and martensitic properties of Fe doped Ni-Mn-In magnetic shape memory alloys have been studied by differential scanning calorimetry, magnetization, resistivity, X-ray diffraction (XRD) and EXAFS. While Ni$_{2}$MnIn$_{1-x}$Fe$_{x}$ ($0 \le x \le 0.6$) alloys are ferromagnetic and non martensitic, the martensitic transformation temperature in Ni$_{2}$Mn$_{1.5} $In$_{1-y}$Fe$_{y}$ and Ni$_{2}$Mn$ _{1.6} $In$_{1-y}$Fe$_{y}$ increases for lower Fe concentrations ($y \le 0.05$) before decreasing sharply for higher Fe concentrations. XRD analysis reveals presence of cubic and tetragonal structural phases in Ni$_{2}$MnIn$_{1-x}$Fe$_{x}$ at room temperature with tetragonal phase content increasing with Fe doping. Even though the local structure around Mn and Ni in these Fe doped alloys is similar to martensitic Mn rich Ni-Mn-In alloys, presence of ferromagnetic interactions and structural disorder induced by Fe affect Mn-Ni-Mn antiferromagnetic interactions resulting in suppression of martensitic transformation in these Fe doped alloys. 
\end{abstract}


\maketitle

\section{Introduction}
The martensitic transformation is one of the most well known magnetostructural property of ferromagnetic shape memory alloys. It's relation to properties like large magnetocaloric effects, magnetic superelasticity \cite{krenke,planes,shamberger} and magnetic field induced giant strains \cite{kainuma,koyama} and exchange bias behaviour\cite{khan} find applications in technology.

In case of Ni$_{2}$MnGa, a relation between the stability of the martensitic state and the dip in the minority-spin density of states (DOS) at the Fermi level leading to a formation of hybrid states of Ni $3d$ and Ga $4p$ minority-spin orbitals has been pointed out \cite{zayak,barman}. Studies on  Ni$_{2+x}$Mn$_{1-x}$Ga have shown that the partial substitution of Mn with Ni results in an increase of martensitic transformation temperature (T$_M$) and a decrease of ferromagnetic ordering temperature (T$_C$) \cite{khovailo,kubler}. EXAFS studies on Ni$_{2+x}$Mn$_{1-x}$Ga have reported the Ni-Ga bond length to be shorter than Ni-Mn bond length resulting in hybridization of Ni $3d$ and Ga $4p$ states \cite{bhobe,sathe}.

Furthermore, local structural disorder has been shown to be an important factor in inducing martensitic transformations in Mn rich Ni$_{2}$Mn$_{1+x}$In$_{1-x}$ alloys \cite{9nelson}. With increasing Mn concentration the Ni-Mn  bond distances decrease much more than Ni-In bond distances giving rise to a local structural disorder which is present well above T$_M$. These local structural distortions are believed to be responsible for Ni $3d$ - Mn $3d$ hybridization near Fermi level. Such a hybridization has been observed through hard X-ray photoelectron spectroscopic studies on Ni$_{2}$Mn$_{1+x}$Sn$_{1-x}$ alloys \cite{9Ye}. This Ni - Mn hybridization also plays an important role in the strength of antiferromagnetic interactions in the martensitic state of these alloys \cite{9nelson3}. These antiferromagnetic interactions are present even in the austenitic phase \cite{9klaer}.

The above studies create an impression that local structural disorder caused by substitution of relatively bigger In atom by a smaller Mn is an important factor responsible for inducing martensitic transformation in Mn rich Ni-Mn-In alloys. The question then arises if martensitic transformation can be induced by doping atoms of similar size as that of Mn. However, doping with Cr or Fe for Mn in Ni$_2$Mn$_{1.4}$In$_{0.6}$ has resulted in very contrasting results. While the substitution of Cr increases the martensitic transition temperature, a similar amount of Fe substitution decreases it drastically \cite{9sharma1}.  Fe substitution is more interesting because very little Fe concentration can suppress martensitic transformation and the suppression of martensitic transformation can be related to strengthening of ferromagnetic interactions.

Fe doping in isostructural Ni$_{2} $MnGa has also shown some interesting trends. On one hand, a complete replacement of Mn with Fe results in T$ _{M} $ to drop from 202K to 145K in Ni$_{2}$MnGa while T$_{C} $ increases from 376K in Ni$ _{2} $MnGa to 430K in Ni$_{2}$FeGa  \cite{9brown,9liu,jqli}. On the other hand, studies by Soto et. al. on the effect of Fe substitution for different atomic species on the structural and magnetic transitions in the  Ni-Mn-Ga alloys have indicated a slightly different picture \cite{9soto}. Substitution of Fe in place of Ni in Ni$_{52.5-x}$Mn$_{23}$Ga$_{24.5}$Fe$_{x}$  (1.2 $ \leq  $x $ \leq $5.5), or Fe for Ga in Ni$_{51.4}$Mn$_{24.5}$Ga$_{24.1-x}$Fe$_{x}$ (0.7 $ \leq $ x $ \leq $ 2.0) results in increase in T$_C$ and a decrease in martensitic and premartensitic transformation temperatures. While Fe substitution for Mn in Ni$ _{51.4} $Mn$ _{25.2-x} $Ga$ _{23.4} $Fe$ _{x} $ shows no appreciable change at either T$ _{M}$ or T$ _{C} $. It appears from their study that the transformation temperatures scale with e/a ratio.

The above studies indicate, that the reasons for changes in martensitic transformation temperature and ferromagnetic ordering temperature on Fe doping in Ni-Mn based magnetic shape memory alloys are far from understood. Important question that still remains unanswered is how a small percentage of Fe doping so drastically affects martensitic transformation as well as the magnetic interactions. Could there be other reasons like antisite disorder,  that have been shown to drastically affect martensitic properties of such alloys?  \cite{nelsonjp, felser} Answering this question is also important in larger context of understanding factors responsible for driving the Mn rich Ni-Mn based Heusler alloys towards martensitic transformation as it would further cement the importance of antiferromagnetic interactions in martensitic transformation in these Mn rich Heusler alloys. To seek answers to these questions a study of structural, magnetic and transport properties of Fe doped Ni-Mn-In alloys has been taken up. In particular we have investigated alloys of the type Ni$_{2}$Mn$ _{1.5} $In$_{0.5-y}$Fe$_{y}$, Ni$_{2}$Mn$ _{1.6} $In$_{0.4-y}$Fe$_{y}$ where $0 \le y \le 0.1$ and  Ni$_{2}$MnIn$_{(1-x)}$Fe$_{x}$ with $0 \le x \le 0.55$. These compositions are chosen because, the Mn rich alloys Ni$_2$Mn$_{1.5}$In$_{0.5}$ and Ni$_2$Mn$_{1.6}$In$_{0.4}$ undergo martensitic transformation in paramagnetic state and therefore do not have a ferromagnetically ordered state while Ni$_2$MnIn is a ferromagnetic alloy that does not undergo a martensitic transformation. It will  therefore be of interest to study the effect of Fe doping on magnetic and martensitic transitions of these alloys. Our studies indicate that Fe doping induces a structural disorder which initially increases T$_M$ but at higher doping concentration leads to formation of Fe rich regions. These regions not only strengthen ferromagnetic interactions but also directly affect the Mn-Ni-Mn antiferromagnetic interactions causing a substantial decrease in T$_M$.

\section{Experimental}
The alloys of above composition were prepared by arc melting the weighed constituents in argon atmosphere followed by encapsulating in a evacuated quartz tube and annealing at 750$^\circ$C for 48 hours and subsequent quenching in ice cold water. The prepared alloys were cut in suitable sizes using a low speed diamond saw. A part of the sample was powdered and reannealed using the same procedure above by covering it in tantalum foil. X-ray diffraction (XRD) patterns on these reannealed powders were recorded at room temperature in the angular range of $20^\circ \leq 2\theta \leq 100^\circ$. Resistivity was measured by standard four probe technique using a closed cycle helium refrigerator in  the temperature range from 10 K to 330K. Magnetization measurements were performed  using a SQUID magnetometer. Here the samples were first cooled from room temperature to 5K in zero applied magnetic field and the data was recorded while warming (ZFC) followed by cooling (FCC) and subsequent warming (FCW) in the same applied field of 100 Oe. Differential Scanning calorimetric (DSC) measurements were carried out on some of the alloys to confirm their martensitic transformation temperature. For this pieces of 3 to 4 mg were crimped in aluminum pan and heated and cooled along with a crimped empty pan at the rate of 3$^\circ$/min in the temperature range of 150K - 700K. EXAFS at Ni K and Mn K edges were recorded at Photon Factory using beamline 12C at room temperature. For EXAFS measurements the samples to be used as absorbers, were ground to a fine powder and uniformly distributed on a scotch tape. These sample coated strips were adjusted in number such that the absorption edge jump gave $\Delta\mu t \le 1$ where $\Delta\mu$ is the change in absorption coefficient at the absorption edge and $t$ is the thickness of the absorber. The incident and transmitted photon energies were simultaneously recorded using gas-ionization chambers as detectors. Measurements were carried out from 300 eV below the edge energy to 1000 eV above it with a 5 eV step in the pre-edge region and 2.5 eV step in the EXAFS region. At each edge, at least three scans were collected to average statistical noise. Data analysis was carried out using IFEFFIT \cite{Newville} in ATHENA and ARTEMIS programs \cite{Ravel}. Here theoretical fitting standards were computed with FEFF6 \cite{Ravel2,Zabinsky}. The data in the k range of 2 to 14 \AA$^{-1}$ and R range of  1 to 3 \AA~ was used for analysis.

\section{Results}

\subsection{Ni$_{2}$Mn$ _{1.5} $In$_{0.5-y}$Fe$_{y}$  and Ni$_{2}$Mn$_{1.6} $In$_{0.4-y}$Fe$_{y}$ Alloys}
The martensitic transformation temperature in Ni$_2$Mn$_{1+x}$In$_{1-x}$ alloys scales with excess Mn concentration or the e/a ratio \cite{planes,9krenke1}. The transformation temperature increases quite rapidly and the alloys with $ x > 0.4$ (e/a ratio $>$ 8) undergo martensitic transformation in a paramagnetic state. In the martensitic state, these alloys order magnetically with dominant antiferromagnetic interactions. While Ni$_2$Mn$_{1.5}$In$_{0.5}$ has T$_M$ = 420K, it increases to 500K in Ni$_2$Mn$_{1.6}$In$_{0.4}$ and further to 620K in Ni$_2$Mn$_{1.7}$In$_{0.3}$. This is quite a rapid increase in T$_M$ for a relatively small change in Mn content. To study the effect of similar amount of Fe doping on martensitic transformation temperature, we have prepared Ni$_{2}$Mn$ _{1.5} $In$_{0.5-y}$Fe$_{y}$  and  Ni$_{2}$Mn$ _{1.6} $In$_{0.4-y}$Fe$_{y}$, $0 \le y \le 0.1$ alloys.

\begin{figure}
 \centering
 \includegraphics[width=\columnwidth]{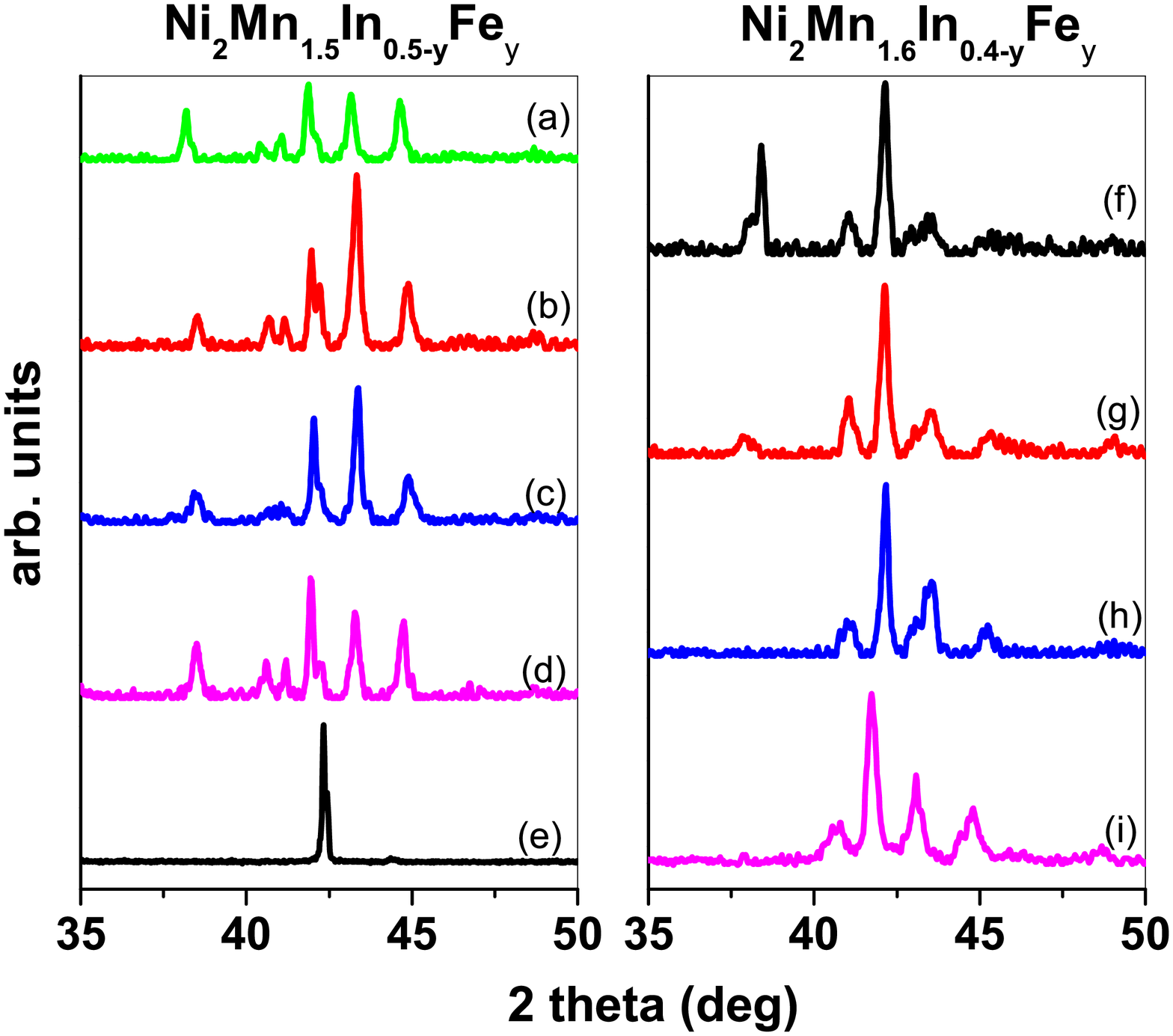}
 \caption{\label{xrd-all.eps}XRD patterns for (a)Ni$_2$Mn$_{1.5}$In$_{0.4}$Fe$_{0.1}$ (b) Ni$_2$Mn$_{1.5}$In$_{0.425}$Fe$_{0.075}$  (c) Ni$_2$Mn$_{1.5}$In$_{0.45}$Fe$_{0.05}$ (d) Ni$_2$Mn$_{1.5}$In$_{0.465}$Fe$_{0.035}$ (e) Ni$_2$Mn$_{1.5}$In$_{0.48}$Fe$_{0.02}$ (f) Ni$_{2}$Mn$ _{1.6} $In$_{0.4}$Fe$_{0.1}$ (g) Ni$_{2}$Mn$ _{1.6} $In$_{0.35}$Fe$_{0.05}$ (h) Ni$_{2}$Mn$ _{1.6} $In$_{0.37}$Fe$_{0.03}$ (i) Ni$_{2}$Mn$ _{1.6} $In$_{0.38}$Fe$_{0.02}$}
 \end{figure}

The XRD patterns for Ni$_{2}$Mn$ _{1.5} $In$_{1-y}$Fe$_{y}$  and  Ni$_{2}$Mn$ _{1.6} $In$_{0.4-y}$Fe$_{y}$ alloys, in limited range of $35 \le 2\theta \le 50$ are presented in Fig \ref{xrd-all.eps}. The XRD patterns in the entire range, $20 \le 2\theta \le 100$ are shown in Supplementary material \cite{supplementary}. The patterns indicate that all the alloys have 7M modulated structure with the exception of Ni$_2$Mn$_{1.5}$In$_{0.4}$Fe$_{0.1}$, which shows a cubic structure at room temperature. This implies that except for Ni$_2$Mn$_{1.5}$In$_{0.4}$Fe$_{0.1}$ all other alloys undergo martensitic transformation above room temperature while Ni$_2$Mn$_{1.5}$In$_{0.4}$Fe$_{0.1}$ either has a stable cubic crystal structure or a martensitic transformation below room temperature.

\begin{figure}
\centering
\includegraphics[width=\columnwidth]{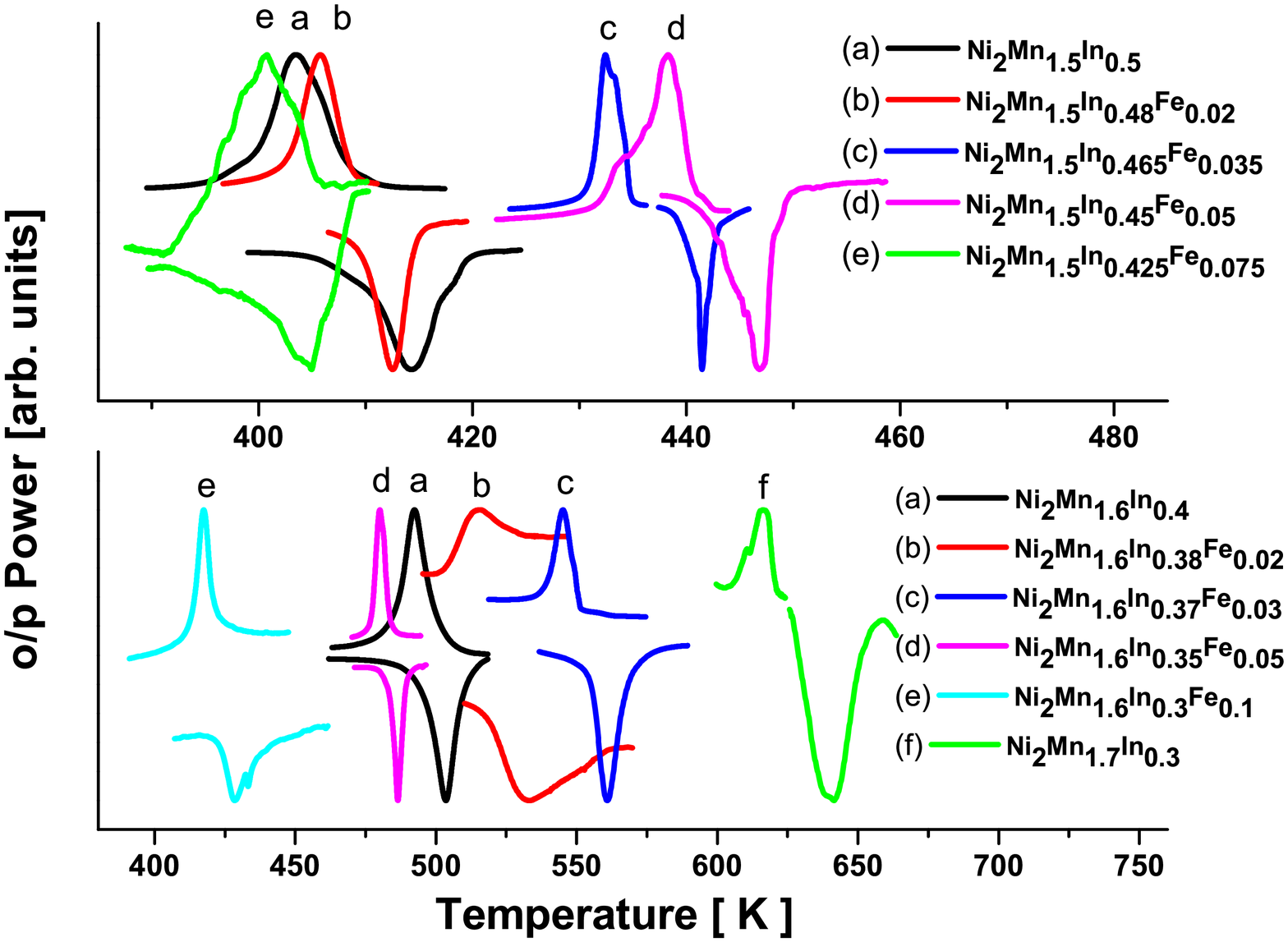}
\caption{\label{dsc-all.eps} DSC plots as a function of temperature for  Ni$_{2}$Mn$ _{1.5} $In$_{0.5-y}$Fe$_{y}$  and Ni$_{2}$Mn$ _{1.6} $In$_{0.4-y}$Fe$_{y}$ alloys }
\end{figure}

In order to determine the martensitic transformation temperatures, DSC measurements were carried out on all the alloys. The heat output curves during heating and cooling cycles are presented  for all alloys except for Ni$_2$Mn$_{1.5}$In$_{0.4}$Fe$_{0.1}$in Fig \ref{dsc-all.eps}. Relatively sharp exothermic and endothermic peaks with hysteretic behaviour were observed respectively during cooling and heating cycles in all samples. These peaks indicate a transformation from high temperature austenitic phase to low temperature martensitic phase. The transformation temperature T$_M$ is then taken as average temperature of exothermic and endothermic peaks.

It can be seen that in Ni$_{2}$Mn$ _{1.5} $In$_{0.5-y}$Fe$_{y}$, T$ _{M} $ increases slowly up to 5\% Fe doping and then decreases rapidly. In the case of Ni$_2$Mn$_{1.5}$In$_{0.4}$Fe$_{0.1}$, the martensitic transformation temperature is well below room temperature (at 120K as deduced from resistivity measurements (not shown)). This fits well to the observed cubic austenitic structure in XRD at room temperature of this alloy. A similar variation of T$_{M}$ is also observed in  Ni$_{2}$Mn$ _{1.6}$In$_{0.4-y}$Fe$_{y}$.

\begin{figure}
 \centering
 \includegraphics[width=\columnwidth]{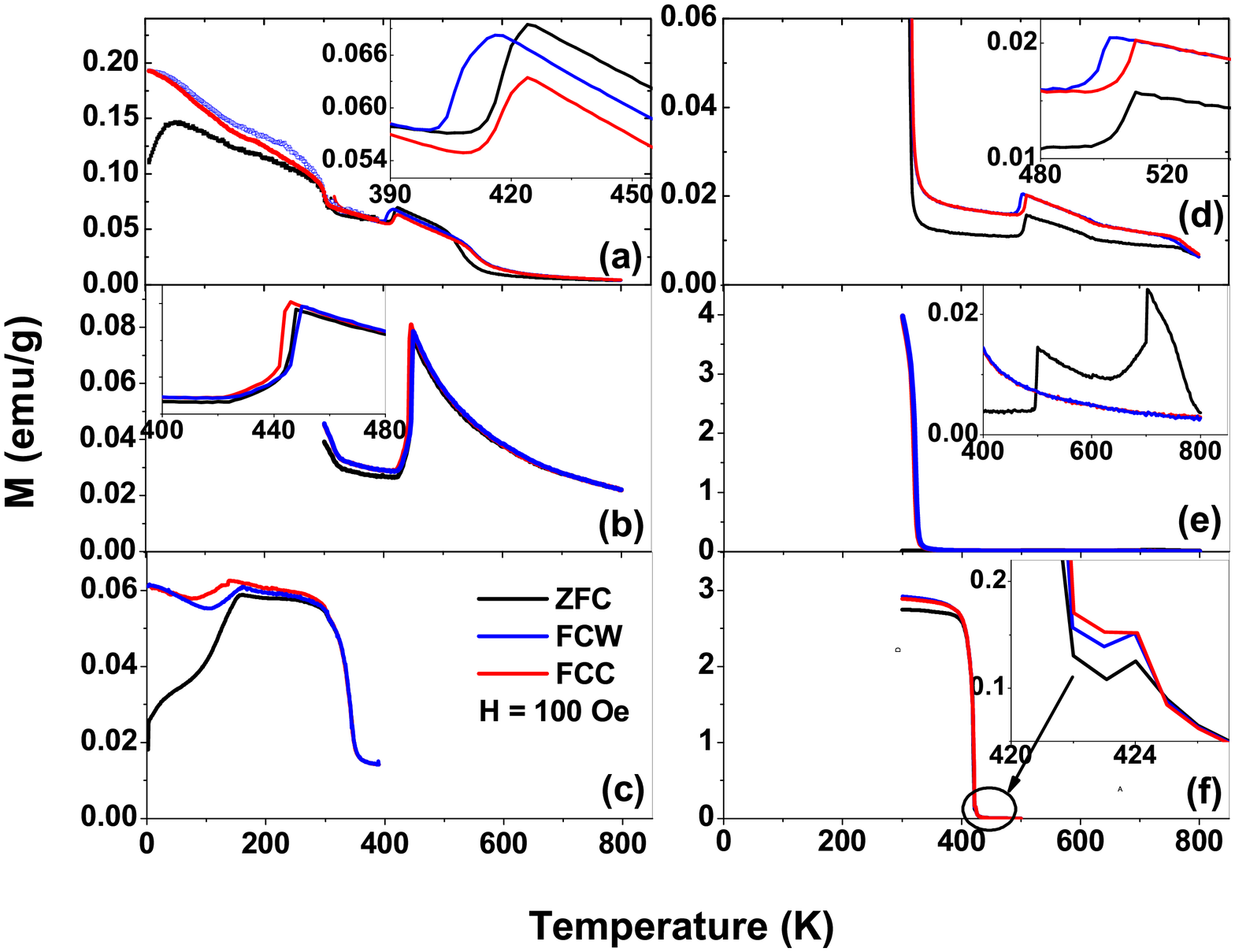}
 \caption{\label{mag-fe1.eps}Magnetization for (a)Ni$_{2}$Mn$ _{1.5} $In$_{0.5}$ (b) Ni$_{2}$Mn$ _{1.5} $In$_{0.45}$Fe$_{0.05}$ (c) Ni$_{2}$Mn$ _{1.5} $In$_{0.4}$Fe$_{0.1}$ (d) Ni$_{2}$Mn$ _{1.6} $In$_{0.4}$ (e) Ni$_{2}$Mn$ _{1.6} $In$_{0.35}$Fe$_{0.05}$ and  (f) Ni$_{2}$Mn$ _{1.6} $In$_{0.3}$Fe$_{0.1}$ alloys}
 \end{figure}

Magnetization measurements have been carried out for y = 0, 0.05 and 0.1 for the alloys Ni$_{2}$Mn$ _{1.5} $In$_{0.5-y}$Fe$_{y}$ and Ni$_{2}$Mn$ _{1.6} $In$_{0.4-y}$Fe$_{y}$ in a field of 100 Oe during ZFC, FCC and FCW cycles and are plotted in FIG \ref{mag-fe1.eps}. It can be seen that, both, Ni$_{2}$Mn$ _{1.5} $In$_{0.5}$ (Fig \ref{mag-fe1.eps}(a)) and Ni$_{2}$Mn$ _{1.6} $In$_{0.4}$ (Fig \ref{mag-fe1.eps}(d)) undergo martensitic transformation in paramagnetic state as indicated by insets in Fig \ref{mag-fe1.eps} (a) and (d). No signature of ferromagnetic transition is observed down to 5K in Ni$_{2}$Mn$ _{1.5} $In$_{0.5}$. Similarly the alloys with $y$ = 0.05 for both series (Fig \ref{mag-fe1.eps} (b) and (e)) also undergo martensitic transformation in paramagnetic state at 420K and 500K respectively. While a sharp rise in magnetization is observed just above 300K in Ni$_{2}$Mn$ _{1.6} $In$_{0.35}$Fe$_{0.05}$ indicating a possibility of magnetic transition, no such transition is observed in Ni$_{2}$Mn$ _{1.5} $In$_{0.45}$Fe$_{0.05}$. However, a small increase in magnetization is observed just above 300K in Ni$_{2}$Mn$_{1.5}$In$_{0.45}$Fe$_{0.05}$ (Fig 3(b)) and may be associated to magnetic ordering in the martensitic phase. A clear paramagnetic to ferromagnetic transitions are observed in $y$ = 0.1 alloys of both series (Fig \ref{mag-fe1.eps} (c) and (f)). Martensitic transformation can also be seen to occur in both the alloys at 120K (Fig \ref{mag-fe1.eps}(c)) and at about 425K (Fig \ref{mag-fe1.eps}(f)). While the values of T$_M$ obtained from magnetization curves agree quite well with those observed from DSC and resistivity, the magnetization measurements distinctly show that not only T$ _{C} $ increases with increase in Fe concentration, ferromagnetism can also be induced in alloys with no prior ferromagnetic transition by Fe doping. 

\begin{figure}
\centering
\includegraphics[width=\columnwidth]{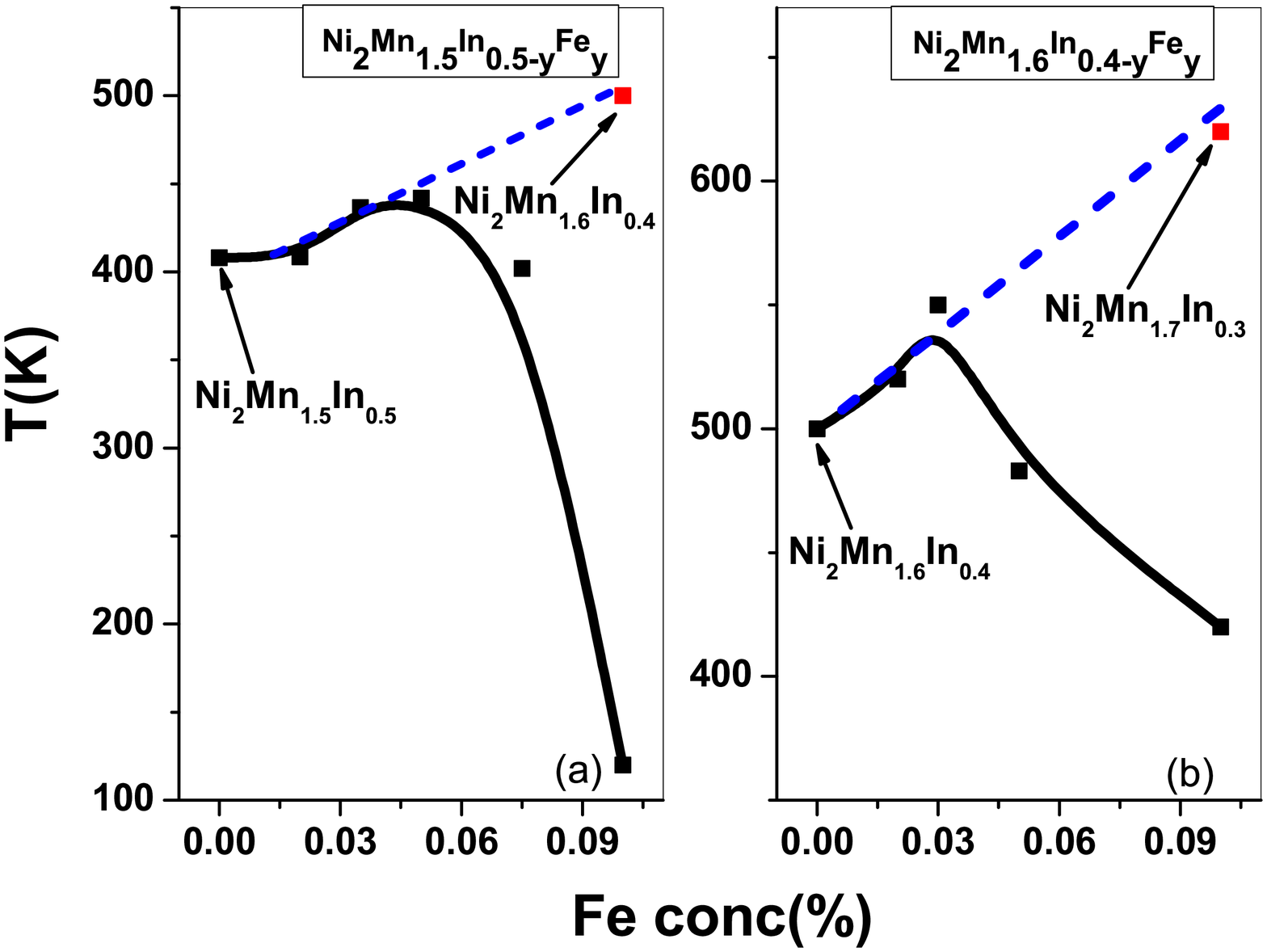}
\caption{\label{Fe-conc.eps}Variation of T$ _{M} $ with Fe concentration for (a) Ni$_{2}$Mn$ _{1.5} $In$_{0.5-y}$Fe$_{y}$ and (b)Ni$_{2}$Mn$ _{1.6} $In$_{0.4-y}$Fe$_{y}$ alloy. }
\end{figure}

The variation of T$ _{M} $ with Fe concentration is plotted in Fig \ref{Fe-conc.eps}. In the case for Ni$_{2}$Mn$ _{1.5} $In$_{0.5-y}$Fe$_{y}$, the undoped compound transforms at 420K. When Mn content is increased to 1.6, T$ _{M} $ increases to 500K as shown by the blue dashed line in Fig \ref{Fe-conc.eps}. As Fe is doped in the alloy to realize  Ni$_{2}$Mn$ _{1.5} $In$_{0.4}$Fe$_{0.1}$, the T$ _{M} $ initially increases to 440K for y = 0.05 and then sharply decreases to 120K for y = 0.1 as depicted by solid black line. Though a similar behaviour is seen in the case of  Ni$_{2}$Mn$ _{1.6} $In$_{0.4-y}$Fe$_{y}$ there is a distinct difference. The rate of decrease of T$_M$ in Ni$_{2}$Mn$ _{1.5} $In$_{0.5-y}$Fe$_{y}$ is much sharper than in Ni$_{2}$Mn$ _{1.6} $In$_{0.4-y}$Fe$_{y}$ alloys. This could be due to higher amount of Mn atoms present at In sites.

\subsection{Ni$_{2}$MnIn$_{1-x}$Fe$_{x}$ Alloys}
\begin{figure}
\centering
\includegraphics[width=\columnwidth]{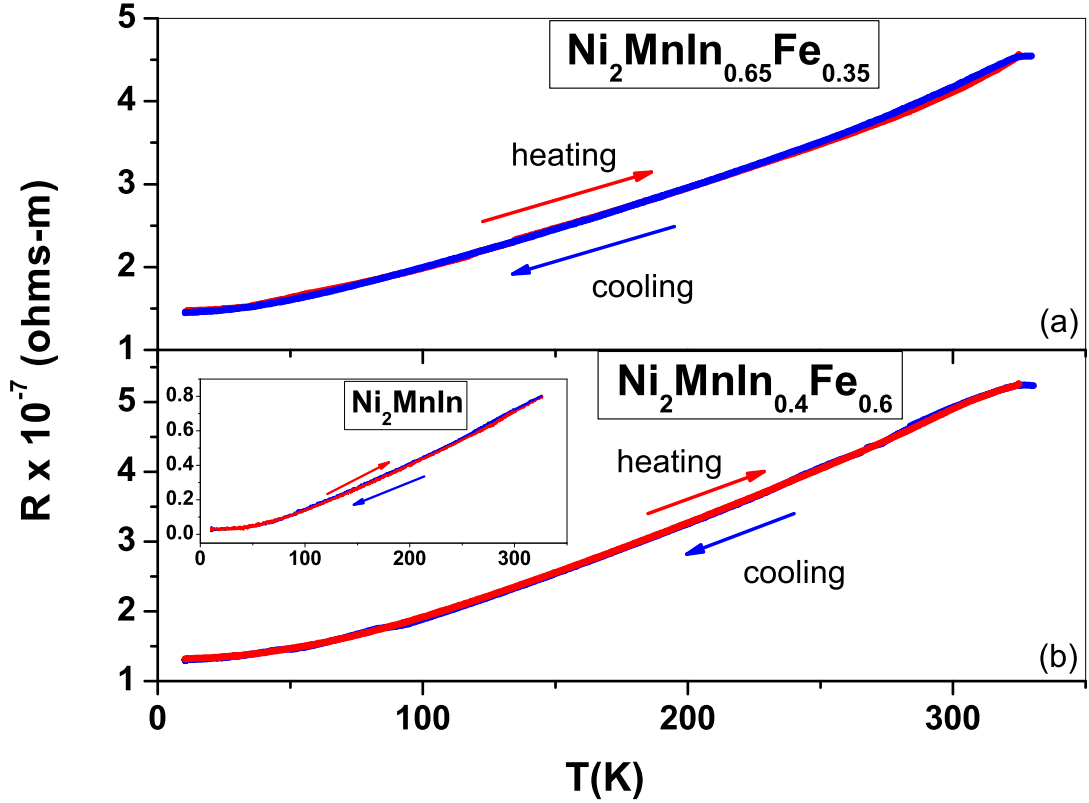}
\caption{\label{res.eps}Resistivity for Ni$_{2}$MnIn$_{(1-x)}$Fe$_{x}$ alloys (a) for x=0.35 and (b)  for x=0.6.
The red and blue lines indicate warming and cooling cycles respectively. The inset of (b) is the resistivity plot for  Ni$ _{2} $MnIn }
\end{figure}

In order to study the effect of Fe substitution in alloys that do not previously undergo martensitic transition, the alloys of the type Ni$_2$MnIn$_{1-x}$Fe$_x$, $0 \le x \le 0.6$, were prepared. The resistivity plots for $x$ = 0.35 and 0.6 are presented in Fig \ref{res.eps}(a) and (b) respectively. These alloys show metallic behaviour throughout the temperature range from 300K to 10K with no discontinuity indicating martensitic transformation. The variation of resistivity is also identical to that in N$_2$MnIn as can be seen in inset of Fig \ref{res.eps}(b). The other alloys of this series also exhibit similar metallic behaviour.

\begin{figure}
\centering
\includegraphics[width=\columnwidth]{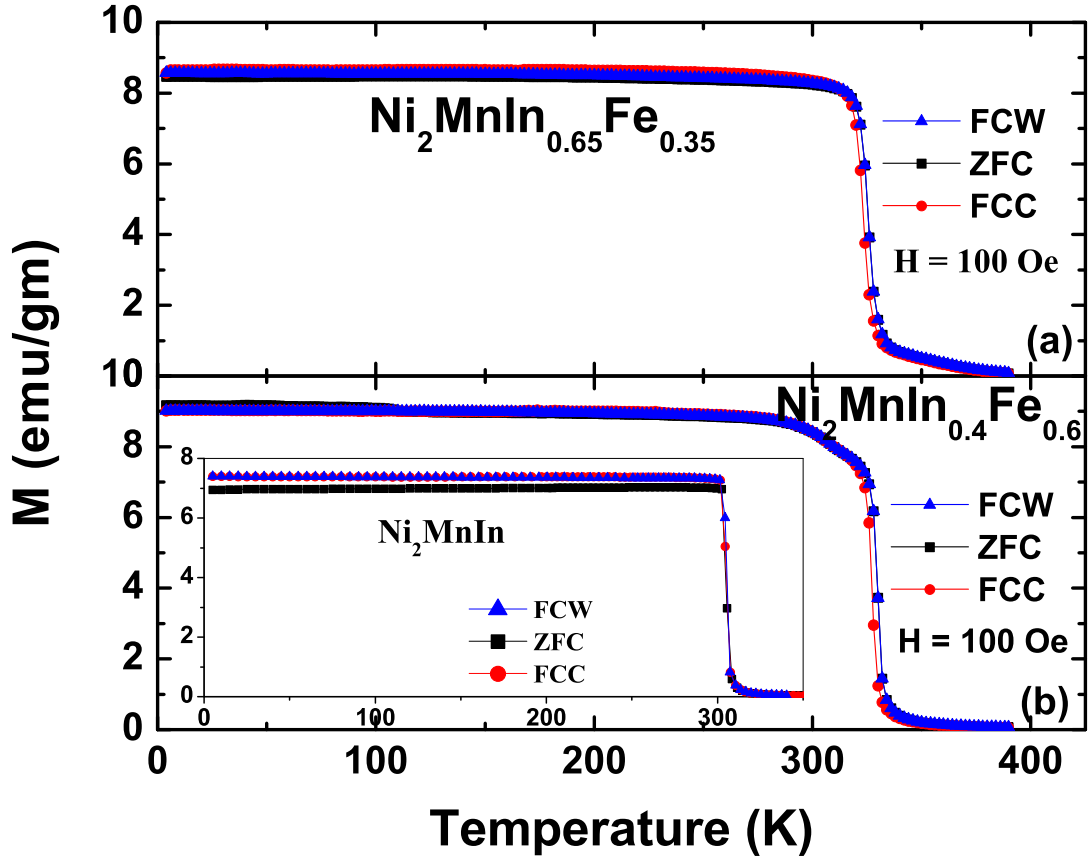}
\caption{\label{mag-fe.eps}Magnetization for Ni$_{2}$MnIn$_{1-x}$Fe$_{x}$ alloys (a) for x = 0.35 and (b)  for x = 0.6}
\end{figure}

Magnetization studies carried out on the same two alloys also support observations drawn from resistivity. Plots of magnetization as a function of temperature for x = 0.35 and  x = 0.6 are plotted in Fig \ref{mag-fe.eps}. The  alloys undergo a paramagnetic to ferromagnetic transition at 333K and 334K respectively and no other transitions are noted down to 5K. Ni$_2$MnIn undergoes paramagnetic to ferromagnetic transition at 306K (inset of Fig \ref{mag-fe.eps}(b)). Therefore there is an increase in T$ _{C} $ due to Fe doping and is also supported by other studies in literature \cite{9liu,9sutou1}. At this stage a common observation that can be drawn is that irrespective of Mn concentration, doping of Fe leads to strengthening of ferromagnetic interactions. However, no martensitic transition is observed in Ni$_2$MnIn$_{1-x}$Fe$_x$ in spite of Fe content being in the same range as that in Ni$_2$Mn$_{1+x}$In$_{1-x}$. This is surprising as Fe should have introduced a similar local structural disorder as Mn upon doping in Ni$_2$MnIn. In order to understand the crystal and local structure of these Fe doped Ni$_2$MnIn we have carried out XRD and Ni and Mn K edge EXAFS studies.

\begin{figure}
\centering
\includegraphics[width=\columnwidth]{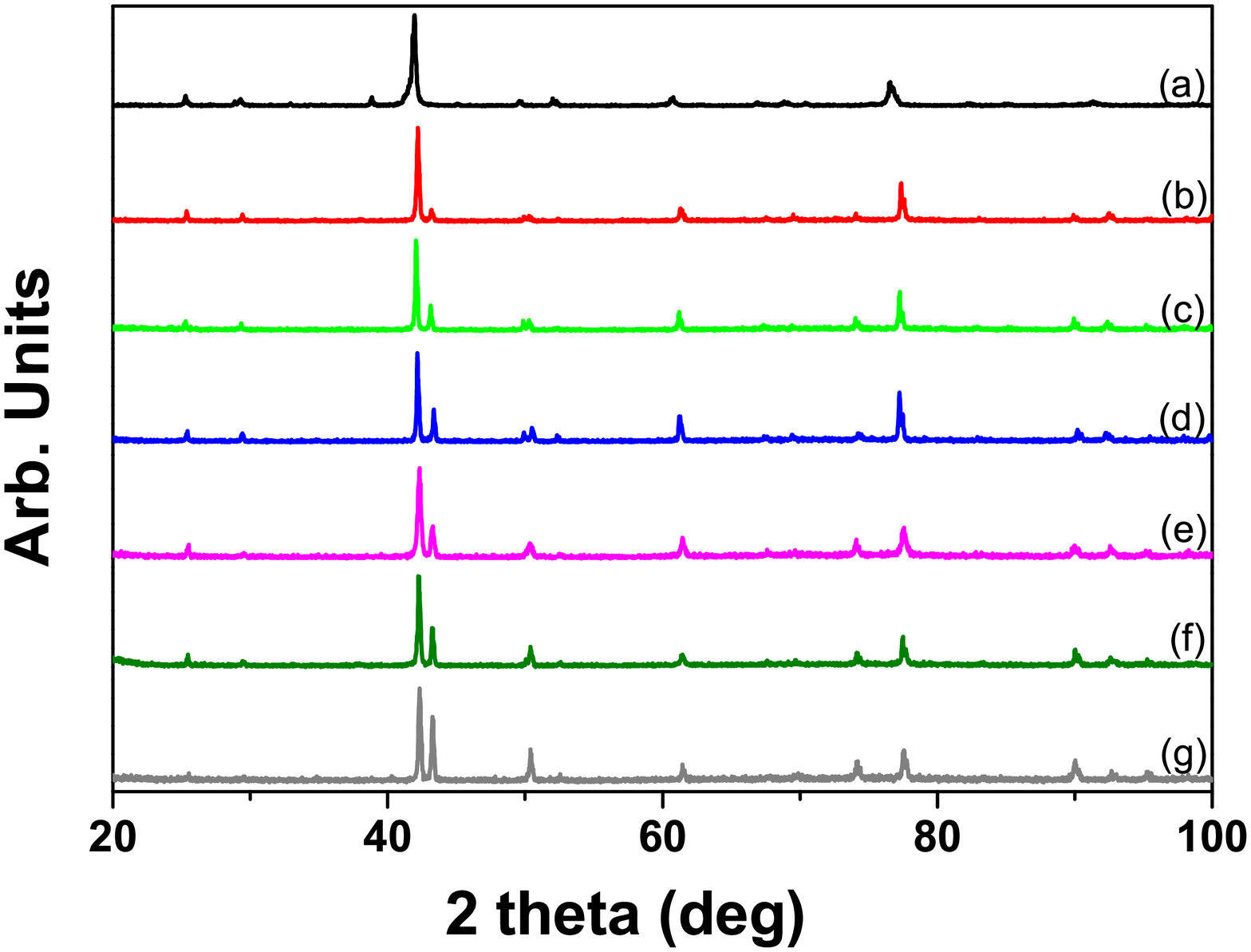}
\caption{\label{Fexrd.eps} XRD pattern for (a) Ni$ _{2} $MnIn and   (b) x = 0.35 (c) x = 0.4 (d) x = 0.45 (e) x = 0.5 (f) x = 0.55 (g) x = 0.6 for  Ni$_{2}$MnIn$_{1-x}$Fe$_{x}$ alloys}
\end{figure}

The room temperature XRD pattern for the series Ni$_{2}$MnIn$_{1-x}$Fe$_{x}$  (0.35 $\leq$ x $\leq$ 0.6) are shown in Fig \ref{Fexrd.eps} and are compared with the room temperature XRD pattern of stiochiometric Ni$ _{2} $MnIn  alloy.  Ni$ _{2} $MnIn is in austenitic L2$ _{1} $ phase of space group Fm\={3}m and a=6.069\AA  \cite{9webster,9krenke1}. For the Fe doped alloys, besides the main (220 )peak at 2$ \theta $ = 42.3 degree, a lower intensity peak at 2$ \theta $ = 43.2 degree is also seen.  This Bragg reflection grows in intensity with increasing Fe doping. The appearance of this along with some other peaks suggests the formation of tetragonal (I4/mmm) phase. The lattice parameters and percentage of cubic and tetragonal phases have been found using Rietveld refinement and are given in Table \ref{ret}. It can be seen that amount of tetragonal phase increases with Fe doping.

The non modulated martensitic alloys have been reported to possess a similar structure however, as seen from magnetization and resistivity studies, these Fe doped alloys do not undergo martensitic transformation. Another possible reason could be the presence of antisite disorder induced by Fe. Fe is known to prefer X as well as Y sites of X$_2$YZ Heusler structure as can be seen by the formation of Fe$_2$MnGa alloys \cite {femn}. Such an antisite disorder can give rise to Fe - Fe ferromagnetic interactions leading to increase in T$_C$.

\begin{table}
\caption{\label{ret}Table of percentage of cubic and tetragonal phases and lattice parameters for Ni$ _{2} $MnIn$ _{1+x} $Fe$ _{x} $ alloys}
\begin{tabular}{|c|c|c|}
\hline
Alloy  & \% composition & lattice parameters \\ \hline
Ni$_2$MnIn$_{0.65}$Fe$_{0.35}$ & tetragonal:13.77(1.10) & a=b= 4.275(1) \\
       & &  c= 5.746(1)    \\
                            & cubic:     86.23 (3.23)    & a = 6.022(0)   \\
Ni$_2$MnIn$_{0.6}$Fe$_{0.4}$ & tetragonal:24.49(1.85)     & a=b= 4.268(0)\\
        & & c= 5.747(1)     \\
                           & cubic:    75.51 (3.80)     & a = 6.023(0)   \\
Ni$_2$MnIn$_{0.55}$Fe$_{0.45}$ & tetragonal:31.03(2.23)     & a=b= 4.271 \\
        & & c= 5.752      \\
                           & cubic:    68.97 (3.87)     & a = 6.019(1)   \\
Ni$_2$MnIn$_{0.5}$Fe$_{0.5}$ & tetragonal:30.70(2.13)     & a=b= 4.273(1) \\
        & & c= 5.761(1)       \\
                            & cubic:     69.30(3.64)     & a = 6.021(1)   \\
Ni$_2$MnIn$_{0.45}$Fe$_{0.55}$ & tetragonal:30.26(2.10)     & a=b= 4.271(0)\\
         & & c= 5.751(0)        \\
                            & cubic:     69.74(3.68)    & a = 6.019(0)   \\
Ni$_2$MnIn$_{0.4}$Fe$_{0.6}$ & tetragonal:38.83(3.04)     & a=b= 4.263(1) \\
         & & c= 5.780(1)        \\
                          & cubic:     61.17(4.26)    & a = 6.020(1)   \\
\hline
\end{tabular}
\end{table}

\begin{figure}
\centering
\includegraphics[width=\columnwidth]{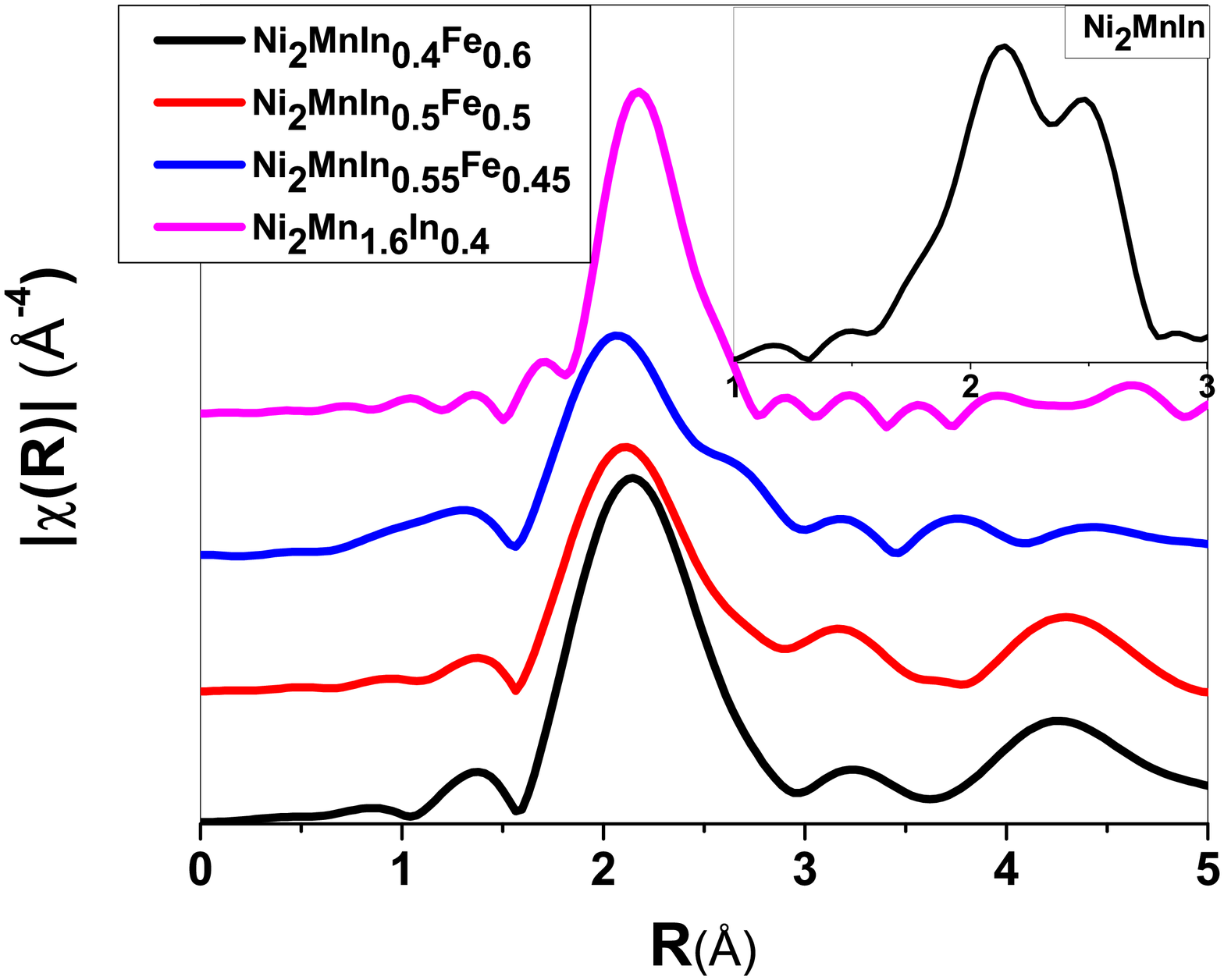}
\caption{\label{athena-fe.eps}Magnitude of FT plots for (a) Ni$_{2}$MnIn$_{(1-x)}$Fe$_{x}$ alloys with inset plot for Ni$ _{2} $MnIn }
\end{figure}

In order to compare the local structures of different atoms in these Fe doped alloys with that of Ni$_{2}$MnIn and Ni$_2$Mn$_{1.4}$In$_{0.6}$ alloys, EXAFS analysis have been carried out at the Ni and Mn K edge for  Ni$_{2}$MnIn$_{1-x}$Fe$_{x}$ alloys. Due to strong absorption by Mn and interference of its EXAFS oscillations, Fe K EXAFS could not be analyzed. A comparison of near edge features at Fe K edge with those of Mn and Ni in the same alloy also did not lead to any clear conclusion. This could either be due to interference of Mn EXAFS and/or site occupancy disorder. The magnitude of Fourier transform (FT) at the Ni K edge  in the interval of R = 1 to 3 \AA \ for x = 0.45, 0.5 and 0.6 along with Ni$_2$Mn$_{1.4}$In$_{0.6}$ are shown in Fig \ref{athena-fe.eps}. The FT magnitude of Ni K EXAFS in Ni$_2$MnIn is also shown as an inset in the Fig \ref{athena-fe.eps}.

The observed FT magnitudes in Fe doped alloys are very similar to that of Ni$_{2}$Mn$_{1.4}$In$_{0.6}$ rather than that in Ni$_2$MnIn. Earlier EXAFS studies \cite{9nelson, 9nelson1} on Ni$_{2}$Mn$_{1+x} $In$_{1-x}$ have reported that the dual peak structure in case of  Ni$_{2}$MnIn is due to different back-scattering amplitudes and phases from equidistant Mn and In atoms surrounding the absorber (Ni) atom. Increasing Mn doping in place of In, results in a single peak due to local structure distortion arising due to shorter Ni-Mn bond distance than Ni-In bond distance \cite{9nelson, 9nelson1}. Therefore, the observation of a single peak in the magnitude of FT plot in Ni$_{2}$MnIn$_{1-x}$Fe$_{x}$ alloys is also suggestive of a local structural disorder around the Ni atom. The various bond lengths obtained for fitting Ni and Mn K edge EXAFS in  Ni$_2$MnIn$_{0.4}$Fe$_{0.6}$ and Ni$_2$MnIn$_{0.5}$Fe$_{0.5}$ alloys are presented in Table \ref{bond}. The fitted spectra in R space and back transformed $k$ space for one of the composition are shown in Supplementary material \cite{supplementary}. The difference in Ni-Mn/Fe and Ni-In bond distances is quite evident from the table. These values of bond distances for the Fe doped alloys compare well with those obtained for Ni$_2$Mn$ _{1.6} $In$ _{0.4}$. Though the observed difference in Ni-Mn/Fe and Ni-In bond distances cannot merely be accounted for by the observed tetragonality in the structure, the increase in the amount of tetragonal phase with increasing Fe content indicates that the local structural disorder evident in EXAFS studies of Ni$_2$MnIn$_{1-x}$Fe$_x$ alloys could be the cause for formation of tetragonal phase.

\begin{table}
\caption{\label{bond}Bond distances obtained from analysis of Ni and Mn K edge EXAFS in Ni$_2$MnIn$_{1+x}$Fe$_{1-x}$ alloys for cubic phase. The numbers in parenthesis indicate uncertainty in the last digit.}   
\begin{tabular}{|l|l|l|l|l|l|l|}
\hline
Sample & \multicolumn{3}{c}{Ni EXAFS} & \multicolumn{3}{|c|}{Mn EXAFS}\\\cline{2-7}
       &\multicolumn{6}{|c|}{Bond length (\AA)}\\\cline{2-7}
       & Ni-Mn/Fe & Ni-In & Ni-Ni & Mn-Ni & Mn-In & Mn-Mn \\
\hline
Ni$_2$Mn$ _{1.6} $In$ _{0.4} $ & 2.557(4) &  2.600(3) &  3.00(1) & 2.574(4) & 3.03(1) & 2.950(1)\\
................. $\sigma^{2}$: & 0.012(1) & 0.010(1) & 0.037(7) & 0.012(1) & 0.029(1) & 0.021(2) \\
Ni$_2$MnIn$_{0.4}$Fe$_{0.6}$   & 2.522(4) & 2.628(4) & 3.02(2) & 2.55(1) & 2.937(6) & 2.925(6) \\
................. $\sigma^{2}$: & 0.014(1) & 0.004(1) & 0.023(2) & 0.012(1) & 0.008(1) & 0.011(1) \\
Ni$_2$MnIn$_{0.5}$Fe$_{0.5}$   & 2.534(4) & 2.622(2) & 3.04(1) & 2.550(3) & 2.931(4) & 2.91(1) \\
.................$\sigma^{2}$: & 0.019(1) & 0.004(1) & 0.022(1) & 0.012(1) & 0.013(1) & 0.012(1) \\
\hline
\end{tabular}
\end{table}

\section{Discussion}
The studies on structural, magnetic and transport properties of Fe doped Ni-Mn-In alloys of the type Ni$_{2}$Mn$_{1.5}$In$_{0.5-y}$Fe$_{y}$, Ni$_{2}$Mn$ _{1.6} $In$_{0.4-y}$Fe$_{y}$ and Ni$_2$MnIn$_{1+x}$Fe$_{1-x}$ indicate that Fe doping, irrespective of Mn content induces ferromagnetic correlations that lead to alloys either ordering ferromagnetically or display an increase in their T$_C$. Further, Fe doping also leads to increased structural disorder, which could be a result of site occupancy disorder. Such a disorder severely affects martensitic transformation temperature. This is clear from the fact that at low Fe doping concentration in martensitic Ni-Mn-In alloys, T$_M$ increases slightly and appears to scale on the same lines as the increase in T$_M$ with Mn doping in Ni$_2$MnIn. However, beyond a certain critical concentration which depends on Mn content, T$_M$ significantly reduces, but at a rate that is dependent on Mn content in the starting composition.

In the case of Ni$_{2}$Mn$_{1.5}$In$_{0.5-y}$Fe$_{y}$, increasing the Fe substitution for In up to 5 \%, causes the martensitic transformation temperature to increase. Such a situation is also observed for  Ni$_{2}$Mn$ _{1.6} $In$_{0.4-y}$Fe$_{y}$ ( $y$ $\leq$ 0.03) alloys. Further increase in Fe content leads to a sharp drop in  martensitic transformation temperature. While in case of Ni$_{2}$Mn$_{1.5}$In$_{0.5-y}$Fe$_{y}$, T$_M$ reduces from 440K ($y$ = 0.05) to 120K ($y$ = 0.1), in Ni$_{2}$Mn$_{1.6}$In$_{0.4-y}$Fe$_{y}$, T$_M$ decreases only to 410K ($y$ = 0.1) from 540K ($y$ = 0.03). This clearly brings out the fact that rate of decrease of T$_M$ depends on the total Mn content.

EXAFS studies on  Ni$_{2}$Mn$ _{1+x} $In$_{1-x}$ have shown the importance of local structure distortions in inducing martensitic transformation \cite{9nelson1}. Such local structural distortions lead to a hybridization of Ni $3d$ and Mn $3d$ orbitals and which plays a role in antiferromagnetic interactions in the martensitic state \cite{9nelson3}. Since Fe is of similar size as that of Mn, substitution of Fe should also cause structural distortions and increase of $T_M$. Such a behaviour is indeed observed for lower Fe concentrations. Increase Fe doping content perhaps also leads to site occupancy disorder which beyond a certain critical concentration of Fe leads to formation of tetragonal phase. This fact can be clearly seen in case of Ni$_2$MnIn$_{1-x}$Fe$_x$. Such a disorder could result in formation of Fe-Fe ferromagnetic interactions thereby increasing $T_C$ or inducing ferromagnetic order. This structural disorder associated with Fe substitution also affects the antiferromagnetic Ni-Mn exchange occurring due to Ni $3d$ - Mn $3d$ hybridization \cite{Sahariah}. This explains the different rates of decrease in T$_M$ for the two series of compounds. Higher the Mn concentration, more are Ni-Mn hybridized pairs and hence slower the  decrease in T$_M$. This compares well with our observations as well as those reported in Ref \cite{9sharma1}. The complete suppression of martensitic behaviour in Ni$_{2}$MnIn$_{1-x}$Fe$_{x}$ alloys is also due to crystallographic phase separation of cubic phase to tetragonal phase with increasing Fe content. This phase separation is a result of structural disorder caused be Fe doping and leads to Fe-Fe ferromagnetic exchange interactions. Therefore, in spite of having e/a ratio similar to Mn rich Ni$_{2}$Mn$_{1+x}$In$_{1-x}$ alloys, the Ni$_{2}$MnIn$_{1-x}$Fe$_{x}$ alloys do not undergo martensitic transformation. In other words, along with presence of local structural distortions, the hybridization of the type [Ni $3d$ - Mn $3d$] at E$ _{F} $ which gives rise to antiferromagnetic exchange interaction could be the important factors that are responsible for martensitic transformation in such Mn rich Ni-Mn based Heusler alloys.

\section{Conclusion}
A systematic study of the martensitic and magnetic interactions in Fe doped  Ni$_{2}$MnIn$_{1-x}$Fe$_{x}$, Ni$_{2}$Mn$ _{1.5} $In$_{0.5-y}$Fe$_{y}$ and Ni$_{2}$Mn$ _{1.6} $In$_{0.4-y}$Fe$_{y}$ Heusler alloys has been carried out. While Ni$_{2}$MnIn$_{1-x}$Fe$_{x}$ (0.35 $\leq$ x $\leq$ 0.6) are non martensitic,  Ni$_{2}$Mn$ _{1.5} $In$_{0.5-y}$Fe$_{y}$ and Ni$_{2}$Mn$ _{1.6} $In$_{0.4-y}$Fe$_{y}$ are martensitic and follow an interesting pattern. For Fe concentrations $ \leq $ 5 \%, the martensitic transformation temperature increases in the same manner as Mn rich Ni-Mn-In alloys and then drops for higher Fe concentrations. This can be attributed to a structural disorder that affects Ni $3d$ - Mn $3d$ hybridization pairs resulting in suppression of martensitic transformation. Larger the affected number of such hybridized pairs, sharper decrease in T$ _{M} $. The same hypothesis can also explain the non martensitic nature of Ni$_{2}$MnIn$_{1-x}$Fe$_{x}$ alloys. Here the structural disorder results in the formation of tetragonal phase which supports direct Fe-Fe ferromagnetic interactions. Therefore, in spite of having similar e/a ratio, these Fe doped alloys are not martensitic. Our results suggest that the local structure disorder and  the Ni $3d$ - Mn $3d$ hybridization at the Fermi level may be responsible factors for martensitic transformation of such Mn rich Ni-Mn based Heusler alloys.

\section*{Acknowledgements}
The work at Photon Factory was performed under the Proposal No. 2011G0077. Financial assistance from CSIR, New Delhi under the project 03(1188)/11/EMR-II and DST, New Delhi is acknowledged.


\begin{thebibliography}{100}
\bibitem{krenke}  T. Krenke, E. Duman, M. Acet, E. F. Wassermann, X. Moya, L. Ma\~nosa and A. Planes, \textit{Phys. Rev B}, \textbf{75} 104414 (2007).
\bibitem{planes} A. Planes, L. Ma\~nosa, and M. Acet, \textit{J. Phys. Condens. Matter}, \textbf{21} 233201 (2009).
\bibitem{shamberger} P. J. Shamberger and F. S. Ohuchi, \textit{Phys. Rev. B} \textbf{79} 144407 (2009).
\bibitem{kainuma}  R. Kainuma, Y. Imano, W. Ito, Y. Sutou, H. Morito, S. Okamoto, O. Kitakami, K. Oikawa, A. Fujita, T. Kanomata and K. Ishida, \textit{Nature(London)} \textbf{439} 957 (2006).
\bibitem{koyama} K. Koyama, K. Watanabe, T. Kaomata, R. Kainuma, K. Oikawa and  K. Ishida, \textit{Appl. Phys. Lett.} \textbf{88} 132506 (2006).
\bibitem{khan} M. Khan, I. Dubenko, S. Stadler and N. Ali, \textit{J. Appl. Phys.} \textbf{102} 113914 (2007).
\bibitem{zayak} A. T. Zayak, P. Entel, K. M. Rabe, W. A. Adeagbo and M. Acet, \textit{ Phys. Rev. B} \textbf{72} 054113 (2005).
\bibitem{barman} S .R. Barman, S. Banik and A. Chakrabarti, \textit{Phys. Rev. B} \textbf{72} 184410 (2005).
\bibitem{khovailo}  V. V. Khovaylo, D. Buchelnikov, R. Kainuma, V. V. Koledov,  M. Ohtsuka, V. G. Shavrov, T. Takagi, S. V. Taskaev and A. N. Vasiliev, \textit{Phys. Rev. B}, \textbf{72} 224408 (2005).
\bibitem{kubler} J. Kubler, A. R. Williams and C. C. Sommers, \textit{Phys. Rev. B}, \textbf{28} 1745 (1983).
\bibitem{bhobe} P. A. Bhobe, K. R. Priolkar and P. R. Sarode, \textit{Phys. Rev. B} \textbf{74} 224425 (2006).
\bibitem{sathe} V. G. Sathe, A. Dubey, S. Banik, S. R. Barman and L. Olivi, \textit{J. Phys. Condens. Matter} \textbf{25} 046001 (2013).
\bibitem{9nelson}  D. N. Lobo, K. R. Priolkar, P. A. Bhobe, D. Krishnamurthy and S. Emura, \textit{Appl. Phys. Lett.} \textbf{96} 232508 (2010).
\bibitem{9Ye} M. Ye, A. Kimura, Y. Miura, M. Shirai, Y. T.Cui, K. Shimada, H. Namatame, M. Taniguchi, S. Ueda, K. Kobayashi, R. Kainuma, T.  Shishido, K. Fukushima and T. Kanomata, \textit{Phys. Rev. Lett.} \textbf{104} 176401 (2010).
\bibitem{9nelson3}  K. R. Priolkar, P. A. Bhobe, D. N. Lobo, S. W. D'Souza, S. R. Barman, A. Chakrabarti, and S. Emura, \textit{Phys. Rev. B} \textbf{87} 144412 (2013).
\bibitem{9klaer} P. Klaer, H. C. Herper, P. Entel, R.  Niemann, L. Schultz, S. F\"{a}hler and H. J. Elmers, \textit{Phys. Rev. B} \textbf{88} 174414 (2013).
\bibitem{9sharma1}  V. K. Sharma, M. K. Chattopadhyay, S. K. Nath, K. J. S. Sokhey, R. Kumar, P. Tiwari and S. B. Roy, \textit{J. Phys. Condens. Matter} \textbf{22} 486007 (2010).
\bibitem{9webster} P. J. Webster, K. R. A. Ziebeck, S. L. Town and M. S. Peak, \textit{Phil. Mag. B} \textbf{49} 295 (1984).
\bibitem{9brown}  P. J. Brown, J. Crangle, T. Kanomata, M. Matsumoto,K. U. Neumann,B. Ouladdiaf, K. R. A. Ziebeck \textbf{J. Phys. Condens. Matter. }\textbf{14} 10159 (2002).
\bibitem{9liu}  Z. H. Liu, M. Zhang M, Y. T. Cui, Y. Q. Zhou, W. H. Wangh, G. H. Wu, X. X. Zhang, and G. Xiao \textit{Appl. Phys. Lett.} \textbf{82} 424 (2003).
\bibitem{jqli}  J. Q. Li, Z. H. Liu, H. C. Yu, M. Zhang M, Y. Q. Zhou, G. H. Wu, \textit{Solid State Comm.} \textbf{126} 323 (2003).
\bibitem{9soto} D. E. Soto, F. A. Hern\'{a}ndez and H. Flores-Z\'{u}\~{n}iga, \textit{Phys. Rev. B} \textbf{77} 184103 (2008).
\bibitem{nelsonjp} D. N. Lobo, S. Dwivedi, C A. daSilva, N. O. Moreno, K. R. Priolkar and A. K. Nigam, \textit{J. Appl. Phys.} \textbf{114} 173910 (2013). 
\bibitem{felser} A. K. Nayak, M. Nicklas, S. Chadov, C. Shekhar, Y. Skourski, J. Winterlik and C. Felser, \textit{Phys. Rev. Lett.}, \textbf{110} 127204 (2013)
\bibitem{Newville}  M. Newville, \textit{J.  Synchrotron Radiat.}, {\bf 8} 322 (2001).
\bibitem{Ravel}  B. Ravel and M. Newville, \textit{J. Synchrotron Radiat.}, {\bf 12} 537 (2005).
\bibitem{Ravel2}  B. Ravel,  \textit{J. Synchrotron Radiat.}, {bf 8} 314 (2001).
\bibitem{Zabinsky}  S. I. Zabinsky, J. J. Rehr, A.  Ankudinov, R. C. Albers and  M. J. Eller, \textit{Phys. Rev. B} {\bf 52} 2995 (1995).
\bibitem{9krenke1}  T. Krenke, M. Acet, E. F. Wassermann, X. Moya, L. Ma\~nosa and A. Planes \textit{Phys. Rev. B} \textbf{73} 174413 (2006).
\bibitem{supplementary} See Supplementary Material as [URL will be inserted by AIP] for XRD and EXAFS data on Fe doped Ni-Mn-In alloys
\bibitem{9sutou1}  D. E. Soto-Parra, X.  Moya, L. Ma\~nosa, A. Planes, H. Flores-Z\'{u}\~{n}iga, F. Alvarado-Hern\'{a}ndez, R. A. Ochoa-Gamboa, J. A. Matutes-Aquinob and D. R\'{\i}os-Jarac, \textit{Phil. Mag.} \textbf{90}(20) 2771 (2010).
\bibitem{femn} W. Zhu, E. K. Liu, L. Feng, X. D. Tang, J. L. Chen, G. H. Wu, H. Y. Liu, F. B. Meng and H. Z. Luo, \textit{Appl. Phys. Lett.} \textbf{95} 222512 (2009). 
\bibitem{9nelson1}  K. R. Priolkar, D. N. Lobo, P. A. Bhobe, S. Emura and A. K.  Nigam, \textit{EPL} \textbf{94} 38006 (2011).
\bibitem{Sahariah}  M. B. Sahariah, S. Ghosh, C. S. Singh, S. Gowtham and R. Pandey, \textit{J. Phys. Condens. Matter} \textbf{25} 025502 (2013).

\end{thebibliography}
\end{document}


\title{\centerline {\em Supplementary material to the paper titled}
{\bf{\noindent Ferromagnetic interactions and martensitic transformation in Fe doped Ni-Mn-In shape memory alloys}}}

\maketitle

\begin{figure}
\centering
\includegraphics[width=\columnwidth]{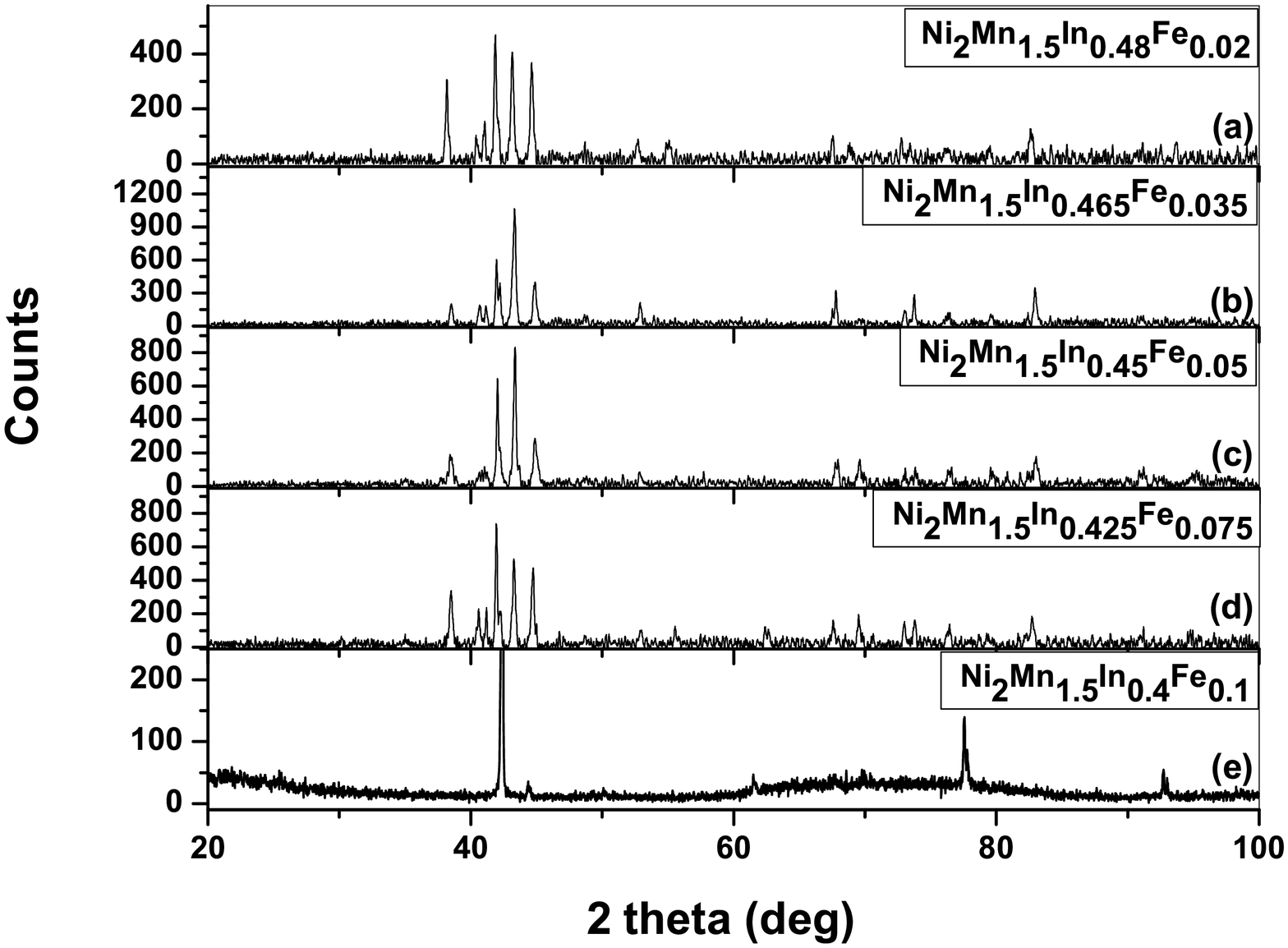}
\includegraphics[width=\columnwidth]{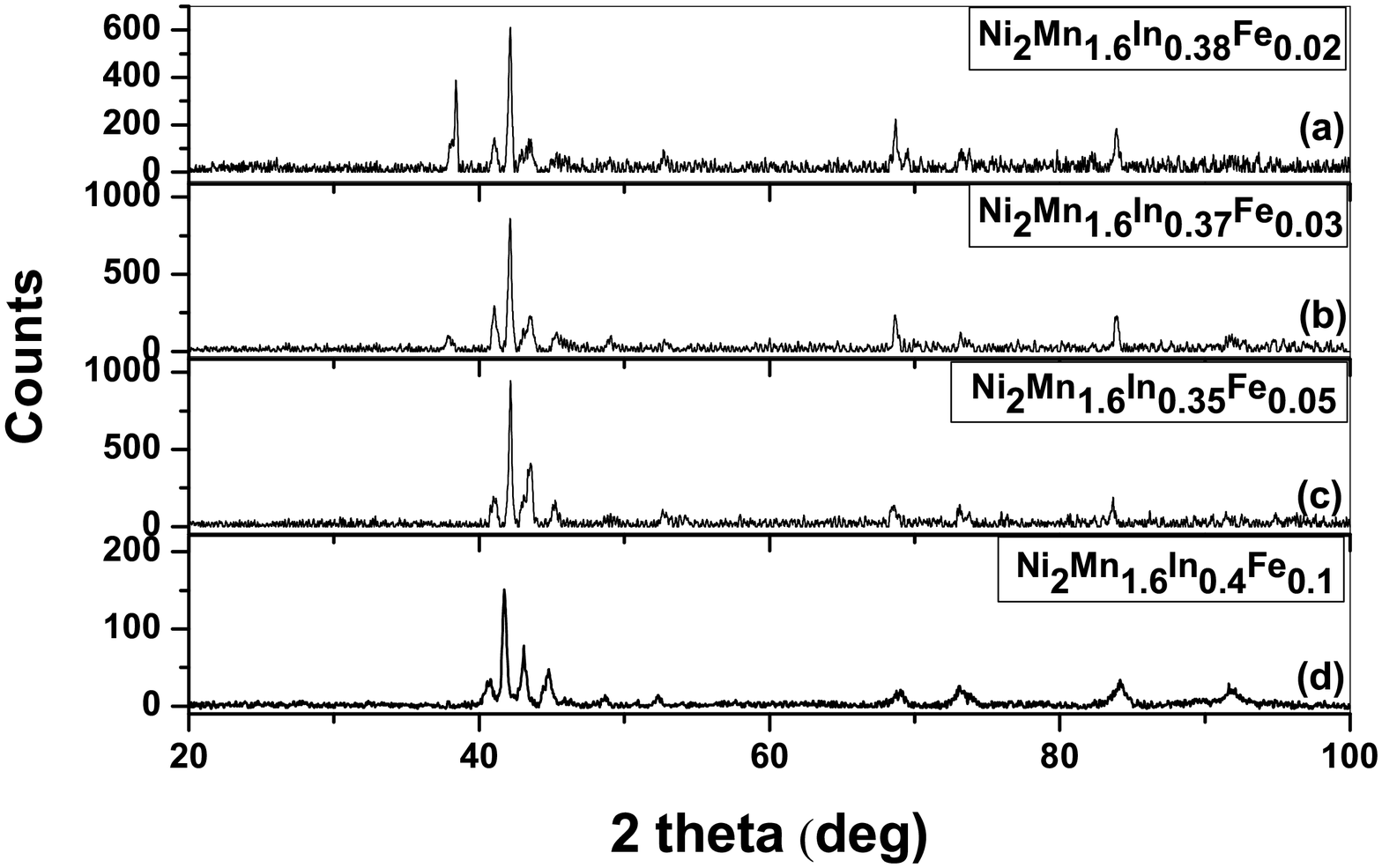}
\caption{\label{mn15} X-ray diffraction patterns for Ni$_{2}$Mn$_{1.5}$In$_{0.5-y}$Fe$_{y}$ and Ni$_{2}$Mn$ _{1.6} $In$_{0.4-y}$Fe$_{y}$ alloys}
\end{figure}

\begin{figure}
\centering
\includegraphics[width=\columnwidth]{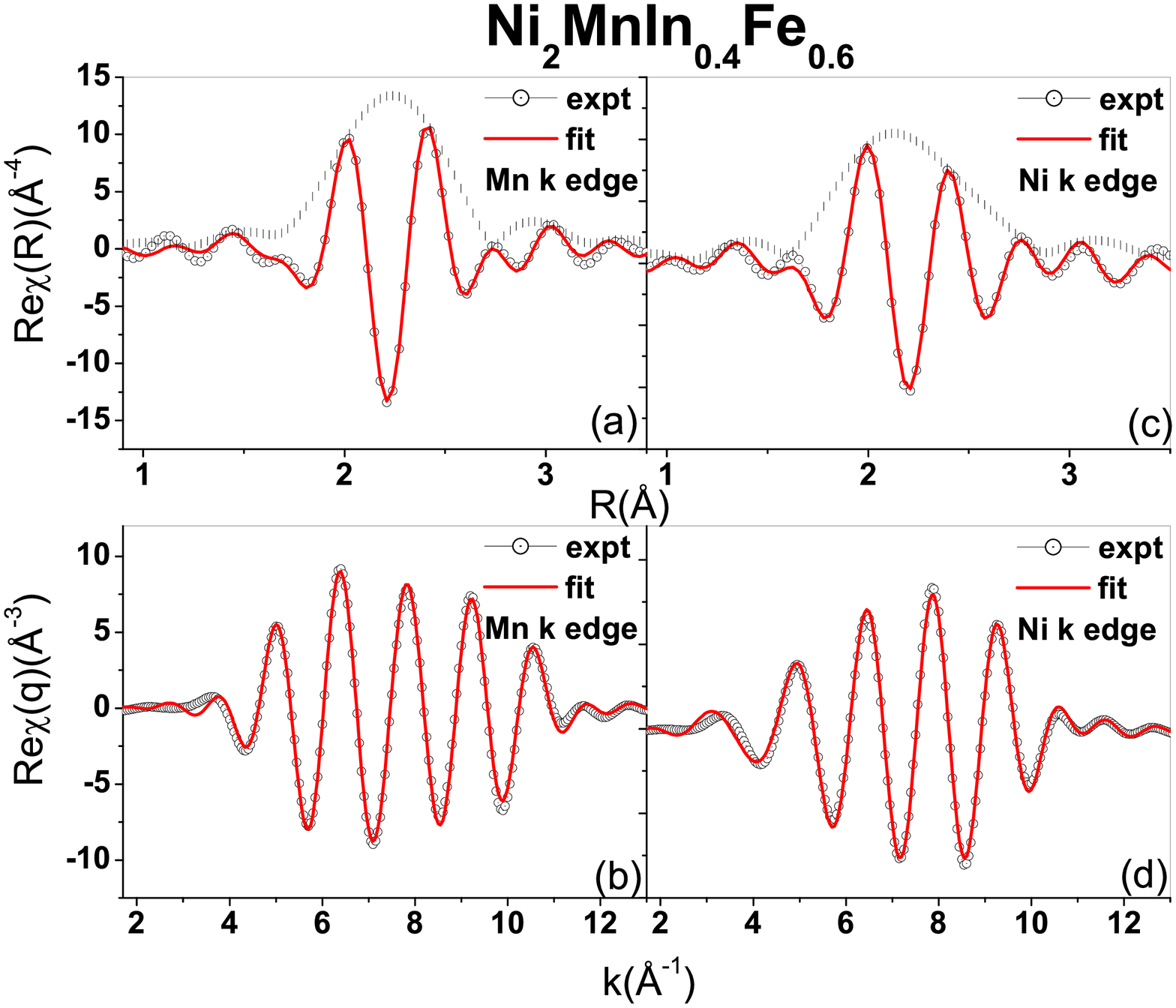}
\caption{\label{Fe6300.eps} (a)and(c) Real component of Mn k edge fourier transform in R and q space (b) and (d) Real component of Ni k edge fourier transform in R and q space for Ni$ _{2} $MnIn$ _{0.4} $Fe$ _{0.6} $}
\end{figure}